\documentclass[11pt,showpacs,
preprintnumbers,amsmath,amssymb,
aps,prd,nofootinbib,eqsecnum]{revtex4}

\pdfoutput=1

\usepackage{epsfig}
\usepackage{graphicx,epsf}
\usepackage{color}
\usepackage{bm}
\usepackage{psfrag}
\usepackage{graphicx}
\usepackage{placeins}
\usepackage{float}
\usepackage[english]{babel}
\usepackage{epstopdf}

\def\be{\begin{equation}}
\def\ee{\end{equation}}
\def\beb{\begin{equation*}}
\def\eeb{\end{equation*}}
\def\bea{\begin{eqnarray}}
\def\eea{\end{eqnarray}}
\def\beab{\begin{eqnarray*}}
\def\eeab{\end{eqnarray*}}
\def\nn{\nonumber}

\def\H{{\cal H}}

\def\1{{}^{(1)}}
\def\2{{}^{(2)}}
\def\cs2{c_{\rm{s}}^2}

\def \beg {\begin{enumerate}}
\def \en {\end{enumerate}}

\def\Pb{P_0}
\def\rhob{\rho_0}

\def\cs{c_{\rm{s}}^2}

\begin{document}

\title{Relating metric and covariant perturbation theories in $f(R)$ gravity}
\author{Adam J.~Christopherson}
\email[]{achristopherson@ufl.edu}
\affiliation{Department of Physics, University of Florida, Gainesville, FL 32611, USA}
\date{\today}

\begin{abstract}
Modified theories of gravity have been invoked recently as an alternative to dark energy, in an attempt
 to explain the apparent accelerated expansion of the universe at the present time. 
 In order to describe inhomogeneities in cosmological models, cosmological perturbation theory
is used, of which two formalisms exist: the metric approach and the covariant approach.
In this paper I present the relationship between the metric and covariant approaches for modeling
 $f(R)$ theories of gravity. This provides a useful resource
that researchers primarily working with one formalism can use to compare or translate their
results to the other formalism.

\end{abstract}


\maketitle
\section{Introduction}

Current observational evidence indicates that we are living in a universe well described 
by $\Lambda$CDM cosmology \cite{WMAP7}. 
This is a model based on the homogeneous and isotropic
Friedmann-Lema\^itre-Robertson-Walker solution to general relativity, complete with
small inhomogeneities, and whose matter content at the present day is dominated by a cosmological constant and
cold dark matter.

However, attempts to reconcile particle physics with general relativity, result in the so-called 
`cosmological constant problem' (see, e.g., Ref.~\cite{Copeland:2006wr}). That is, the 
observed value of $\Lambda$ differs from that predicted from 
fundamental theories by over a hundred orders of magnitude.
In light of this, there has been much recent work 
on attempting to pinpoint the nature of this dark energy component of the universe.
There are essentially two methods of modeling the present day acceleration of the expansion of the universe,
aside from a cosmological constant term. We can either introduce a dark energy fluid with a negative pressure 
(e.g. Ref.~\cite{Copeland:2006wr}), or we can drop the assumption that Einstein's gravity is valid on all scales, adding modifications on the largest scales (e.g. Ref.~\cite{Clifton:2011jh}).

One particular theory of modified gravity consists of a modification of the 
Einstein-Hilbert action to depend upon a function of the Ricci scalar. This is perhaps the most
popular modified gravity theory of recent years, and is called $f(R)$ gravity \cite{Capozziello:2002rd, Nojiri:2003ft, Carroll:2003wy, DeFelice:2010aj}.
This is the modified theory of gravity on which we focus in the present article.

Observational evidence points towards the existence of small inhomogeneities, generated during the inflationary
phase, as the seeds of large scale structure. There are two popular techniques for modeling these inhomogeneities.
The first is inspired by the pioneering early research by Lifshitz \cite{Lifshitz:1945du} and developed by Bardeen \cite{Bardeen:1980kt} and is based around
considering perturbations to the FLRW metric. The second method, dubbed the `covariant approach' follows
work by Ellis, Bruni and collaborators \cite{Ellis:1989jt}. To date, the majority of the study of inhomogeneous perturbations of 
$f(R)$ theories have been completed in the metric formalism (see Ref.~\cite{DeFelice:2010aj} for a detailed 
reference list). However, recently some authors have used the covariant approach to study the perturbations 
\cite{Carloni:2007yv}. These approaches each suffer from their own strengths and weaknesses
(see, e.g., Ref.~\cite{Malik:2012dr}) however, they are equivalent, describing the same physical universe,
 and thus results obtained in each formalism should be in agreement. 

Previous work has studied the equivalence between the two formalisms for Einstein gravity \cite{Bruni1992}. However, to date, the relationship between the formalisms has not been presented for
modified gravity and, in particular, for $f(R)$ gravity. In this paper we perform such a study, extending
Ref.~\cite{Bruni1992} to $f(R)$ gravity. This will enable authors working in one 
formalisms to compare their results to the other. The paper is organized as follows: in the next section we 
review the basics of $f(R)$ gravity and define our notation. In Section \ref{sec:metric} we present perturbations
in the metric formalism, picking two particularly popular gauges before considering the covariant
formalism in Section \ref{sec:cov}. In Section \ref{sec:relation} we relate the two approaches, showing how to 
transform from one to the other, before concluding in Section \ref{sec:dis}.

\section{Basics of $f(R)$ gravity}

In $f(R)$ gravity the Einstein-Hilbert action is modified to take the form 
\be 
S=\frac{1}{2}\int d^4x\sqrt{-g}f(R)+\mathcal{L}_M\,,
\ee
which, on varying with respect to the metric, gives the field equations
\be 
\tilde{G}_{ab}\equiv F(R)R_{ab}-\frac{1}{2}f(R)g_{ab}-\nabla_b\nabla_a F+g_{ab}\Box F=8\pi G T^{M}_{ab}\,,
\ee
where $F\equiv \partial f / \partial R$, $\Box\equiv\nabla_c\nabla^c$, and $T^{M}_{ab}$ is the energy-momentum
tensor, obtained by varying the matter action with respect to the metric. Here, and throughout, 
indices $a,b,\ldots$ cover the full spacetime range $(0,\ldots, 3)$, and $i,j,\ldots$ denote spatial indices $(1,\ldots, 3)$.

Alternatively, one can write the field equations as 
\be 
F G_{ab}=T^{M}_{ab}+\frac{1}{2}g_{ab}(f-RF)+\nabla_b\nabla_a F -g_{ab}\Box F\,,
\ee
where $G_{ab}=R_{ab}-\frac{1}{2}g_{ab}R$ is the usual Einstein tensor. This equation can then
be written as 
\be 
G_{ab}=\tilde{T}^{M}_{ab}+T^{R}_{ab}\,,
\ee
where $\tilde{T}^{M}_{ab}=T^{M}_{ab} / F$ is the rescaled matter energy-momentum tensor, and 
\be 
T^{R}_{ab}=\frac{1}{F}\Big[\frac{1}{2}g_{ab}(f-RF)+\nabla_b\nabla_a f - g_{ab}\Box F\Big]\,,
\ee
is the energy-momentum tensor of the effective `curvature fluid'. 
It is important to be able to write the system as 
general relativity with effective fluids, since it enables one to apply the covariant approach to perturbation theory
to the model.\\

\section{Metric perturbation theory}
\label{sec:metric}

Metric perturbation theory has been studied by many authors over the past 
few decades \cite{ks, mfb} building upon the first comprehensive work on gauge invariant
linear cosmological perturbations conducted by Bardeen \cite{Bardeen:1980kt}. In the years
since, metric perturbation theory has been extended to second order and beyond (see, e.g., 
Refs.~\cite{MW2008, MM2008, thesis, Christopherson:2011hn} 
and references therein), and recently studies have been developed to encompass modified gravity theories, 
such as $f(R)$ (see, e.g., Refs.~\cite{Tsujikawa:2007tg, Hwang:2001qk}).

In metric perturbation theory we consider small, inhomogeneous perturbations to the homogeneous and isotropic
Friedmann-Lema\^itre-Robertson-Walker (FLRW) background spacetime. Doing so gives the perturbed line element
\be 
ds^2=a^2(\eta)\Big[-1(1+2\phi)d\eta^2+2B_{i}dx^i d\eta+(\gamma_{ij}+2C_{ij})dx^idx^j\Big]\,,
\ee
where $\eta$ is conformal time, $a(\eta)$ is the scale factor,
 $\gamma_{ij}$ is the metric on the spatial 3-hypersurface, and a bar denotes the covariant derivative with respect to
this metric. The perturbations can be further split up using the scalar-vector-tensor decomposition \cite{stewart1990} as
\begin{align}
B_{i}&=B_{|i}-S_i\,,\\
C_{ij}&=-\psi\gamma_{ij}+E_{|ij}+F_{(i|j)}+\frac{1}{2}h_{ij}\,.
\end{align}
In Appendix \ref{app:notation} we show the relationship between variables using different notational conventions.

Considering now only scalar perturbations, with a flat spatial metric, results in the 
line element
\be 
ds^2=a^2(\eta)\Big[-(1+2\phi)d\eta^2+2B_{,i}d\eta dx^i+\Big((1-2\psi)\delta_{ij}+2E_{,ij}\Big)dx^idx^j\Big]\,,
\ee
The background
Einstein equations are then
\begin{align}
3F\H^2&=\frac{a^2}{2}(FR-f)-3\H \dot{F}+\rhob\,,\\
-2\dot{\H} F&=\ddot{F}-2\H \dot{F}+a^2(\rhob+\Pb)\,,
\end{align}
where a dot denotes a derivative with respect to conformal time, $\mathcal{H}=\dot{a}/a$ is the Hubble parameter, and the background Ricci scalar is
\be 
R=\frac{6}{a^2}(\H^2+\dot{\H})\,.
\ee

The Einstein equations for the linear perturbations then give an equation from the (ADM) energy constraint
\be 
-\nabla^2\psi+3\H(\dot{\psi}+\H\phi)-\nabla^2(\dot{E}-B)
=\frac{1}{2F}\Big[\dot{F}\nabla^2(\dot{E}-B)-(\nabla^2+3\dot{\H})\delta F+3\H\dot{\delta F}
-3\dot{F}(\dot{\psi}+2\H\phi)-8\pi Ga^2\delta\rho\Big]\,,
\ee 
and from the momentum constraint
\be 
\H\phi+\dot{\psi}=\frac{1}{2F}\Big[\dot{\delta F}-\dot{F}\phi-\H\delta F-8\pi G(\rhob+\Pb)a^2(v+B)\Big]\,,
\ee
where $\delta F = \frac{\partial F}{\partial R} \delta R=F'\delta R$.
From the ADM propagation equation ($\tilde{G}^i{}_j-\frac{1}{3}\delta^i{}_j \tilde{G}^k{}_k$ component),
 after applying the operator $\partial_i\partial^j$,
\be 
\ddot{E}-\dot{B}+2\H(\dot{E}-B)+\psi-\phi = \frac{1}{F}\Big[\delta F-\dot{F}(\dot{E}-B)\Big]\,.
\ee
The Raychaudhuri equation ($\tilde{G}^k{}_k-\tilde{G}^0{}_0$ component) gives the equation
\begin{align}
3\ddot{\psi}+3\H(\dot{\phi}+\dot{\psi})-\nabla^2(\ddot{E}-\dot{B})-2\H\nabla^2(\dot{E}-B)
+\Big(6\dot{\H}+3\H^2+\nabla^2+3\frac{\ddot{F}}{F}\Big)\phi\nn\\
+\frac{\dot{F}}{2F}\Big[3\dot{\psi}+3\dot{\phi}-\nabla^2(\dot{E}-B)\Big] 
=\frac{1}{2F}\Big[3\ddot{\delta F}-(6\H^2+\nabla^2)\delta F-8\pi G a^2(\delta\rho+\delta P)\Big]\,,
\end{align}
and finally, the trace equation ($\tilde{G}^a{}_a\equiv\tilde{G}^0{}_0+\tilde{G}^k{}_k$ component) gives the equation
\be 
\ddot{\delta F}+2\H\dot{\delta F}-\Big(\frac{R}{3}+\nabla^2\Big)\delta F=\frac{8\pi G}{3}a^2(3\delta P-\delta\rho)
+\dot{F}\Big[4\H\dot{\phi}+3\dot{\psi}-\nabla^2(\dot{E}-B)+\dot{\phi}\Big]+2\ddot{F}\phi-\frac{1}{3}a^2F\delta R\,.
\ee

The perturbed Ricci scalar is 
\be 
\delta R = \frac{2}{a^2}\nabla^2\Big(2\psi-\phi+\ddot{E}-\dot{B}\Big)-\frac{6}{a^2}\Big[\H(\dot{\phi}+3\dot{\psi})
-\H\nabla^2(\dot{E}-B)+\ddot{\psi}+2(\dot{\H}+\H^2)\phi\Big]\,,
\ee
where the spatial Laplacian is $\nabla^2\equiv\partial_k\partial^k$.

\subsection{Gauge Choice}

When using cosmological perturbation theory, one encounters the `problem' of gauge invariance.
As described above, the formalism requires the splitting of the
spacetime into a background spacetime and a perturbed spacetime. However, this
method of splitting is not a covariant process.
 That is, one can make a choice of `gauge' which relates points on the background
spacetime to points on the perturbed spacetime, but the choice is not unique. Therefore, quantities can change
depending on the choice of coordinate correspondence.

One resolution of this issue was proposed by Bardeen in 1980 \cite{Bardeen:1980kt} where he first introduced the idea of looking at solely gauge 
invariant variables. These are quantities  constructed such that they do not change under a gauge transformation. This is
equivalent to eliminating the gauge degrees of freedom from the metric from the outset, therefore
guaranteeing that one is working with only gauge invariant variables. In the previous section equations were presented without fixing a gauge. Now, we highlight a couple of common
gauges, and present the governing equations for $f(R)$ gravity theories in these gauges.

\subsubsection{Longitudinal gauge}
\label{sec:long}

The longitudinal gauge is the gauge in which the shear metric perturbation, $\sigma\equiv \dot{E}-B$, is zero. This gives $B=0=E$. The two 
remaining scalar metric perturbations are then the Bardeen potentials \citep{Bardeen:1980kt} defined as
\begin{align}
\Phi&=\phi-\H(\dot{E}-B)-(\ddot{E}-\dot B)\,,\\
\Psi&=\psi+\H(\dot{E}-B)\,,
\end{align}
(or, in Bardeen's notation, $\Phi_A Q^{(0)}$ and $-\Phi_H Q^{(0)}$). The metric then has no off-diagonal 
terms and is \footnote{Note that this is a gauge which does not exhibit problems when transforming between
frames in modified gravity theories. We do not explore this, but see Ref.~\cite{Brown:2011eh} for details.}
\be 
ds^2=a^2(\eta)\Big[-(1+2\Phi)d\eta^2+(1-2\Psi)\delta_{ij}dx^i dx^j\Big]\,.
\ee

The governing equations for linear perturbations in this gauge are then:
\be 
-\nabla^2\Psi+3\H(\dot{\Psi}+\H\Phi)
=\frac{1}{2F}\Big[-(\nabla^2+3\dot{\H})\delta F+3\H\dot{\delta F}
-3\dot{F}(\dot{\Psi}+2\H\Phi)-8\pi Ga^2\delta\rho_\ell\Big]\,,
\ee 
\be 
\H\Phi+\dot{\Psi}=\frac{1}{2F}\Big[\dot{\delta F}-\dot{F}\Phi-\H\delta F-8\pi G(\rhob+\Pb)a^2v_\ell\Big]\,,
\ee
\be 
\Psi-\Phi = \frac{\delta F}{F}\,,
\ee
\begin{align}
3\ddot{\Psi}+3\H(\dot{\Phi}+\dot{\Psi})
&+\Big(6\dot{\H}+3\H^2+\nabla^2+3\frac{\ddot{F}}{F}\Big)\Phi
+\frac{\dot{F}}{2F}\Big[3\dot{\Psi}+3\dot{\Phi}\Big] \nn\\
&=\frac{1}{2F}\Big[3\ddot{\delta F}-(6\H^2+\nabla^2)\delta F-8\pi G a^2(\delta\rho_\ell+\delta P_\ell)\Big]\,,
\end{align}
\be 
\ddot{\delta F}+2\H\dot{\delta F}-\Big(\frac{R}{3}+\nabla^2\Big)\delta F
=\frac{8\pi G}{3}a^2(3\delta P_\ell-\delta\rho_\ell)
+\dot{F}\Big[4\H\dot{\Phi}+3\dot{\Psi}+\dot{\Phi}\Big]+2\ddot{F}\Phi-\frac{1}{3}a^2F\delta R_\ell\,,
\ee
and the Ricci scalar is
\be 
\delta R_\ell = \frac{2}{a^2}\nabla^2\Big(2\Psi-\Phi\Big)-\frac{6}{a^2}\Big[\H(\dot{\Phi}+3\dot{\Psi})
+\ddot{\Psi}+2(\dot{\H}+\H^2)\Phi\Big]\,.
\ee

In Bardeen's original work, the chosen gauge invariant matter variable was the density perturbation in the comoving
gauge. This can be related to the longitudinal gauge variables used above through
\be 
\delta\rho_{\rm com}=\delta\rho_\ell+\dot{\rho_0}v_\ell\,.
\ee

\subsubsection{Uniform curvature gauge}

The uniform curvature gauge is the one in which $E=\psi=0$, and so the metric tensor is then
spatially unperturbed:
\be 
ds^2=a^2(\eta)\Big[-(1+2\phi)d\eta^2+2B_{,i}d\eta dx^i+d{\bm x}^2\Big]\,.
\ee
The governing equations in this gauge are then
\be 
3\H^2\phi+\nabla^2 B
=\frac{1}{2F}\Big[-\dot{F}\nabla^2 B-(\nabla^2+3\dot{\H})\delta F+3\H\dot{\delta F}
-6\dot{F}\H\phi-8\pi Ga^2\delta\rho\Big]\,,
\ee 
\be 
\H\phi=\frac{1}{2F}\Big[\dot{\delta F}-\dot{F}\phi-\H\delta F-8\pi G(\rhob+\Pb)a^2(v+B)\Big]\,.
\ee
\be 
\dot{B}+2\H B+\phi = -\frac{1}{F}\Big[\delta F+\dot{F}B\Big]\,,
\ee
\begin{align}
3\H \dot{\phi}+\nabla^2 \dot{B}+2\H\nabla^2 B
&+\Big(6\dot{\H}+3\H^2+\nabla^2+3\frac{\ddot{F}}{F}\Big)\phi 
+\frac{\dot{F}}{2F}\Big[3\dot{\phi}+\nabla^2 B\Big] \nn\\
&=\frac{1}{2F}\Big[3\ddot{\delta F}-(6\H^2+\nabla^2)\delta F-8\pi G a^2(\delta\rho+\delta P)\Big]\,,
\end{align}
\be 
\ddot{\delta F}+2\H\dot{\delta F}-\Big(\frac{R}{3}+\nabla^2\Big)\delta F=\frac{8\pi G}{3}a^2(3\delta P-\delta\rho)
+\dot{F}\Big[4\H\dot{\phi}+\nabla^2 B+\dot{\phi}\Big]+2\ddot{F}\phi-\frac{1}{3}a^2F\delta R\,,
\ee
while the perturbed Ricci scalar is 
\be 
\delta R = -\frac{2}{a^2}\nabla^2\Big(\phi+\dot{B}\Big)-\frac{6}{a^2}\Big[\H \dot{\phi}
+\H\nabla^2 B+2(\dot{\H}+\H^2)\phi\Big]\,.
\ee

\section{Covariant Formalism}
\label{sec:cov}

The starting point for the covariant approach to cosmological perturbations 
is choosing a suitable frame in which to work. Equivalently, this means making
a choice of the four velocity vector, $u_a$, of an observer in the spacetime. Several different choices can be made, 
but the most
physically motivated choice  is the frame associated with standard matter, so $u_a=u_a^{M}$. Now following closely
Refs.~\cite{Ellis:1998ct, Carloni:2007yv}, we can derive the kinematic quantities in the standard way. In the following we
denote the derivative along the matter fluid flow lines with a dagger, e.g., ${X}^\dagger=u_a\nabla^a X$. 

The projection tensor is 
\be 
h_{ab}\equiv g_{ab}+u_a u_b\,,
\ee
which obeys
\be 
h^a{}_b h^b{}_c = h^a {}_c\,, \, \, \, 
h_{ab}u^b=0\,.
\ee
The projected derivative operator orthogonal to $u^a$ is ${}^{(3)}{\nabla_a}=h^b{}_a\nabla_b$, and so kinematical
quantities are introduced by splitting the covariant derivative of $u^a$:
\be 
\nabla_b u_a = {}^{(3)}{\nabla}_b u_a - a_a u_b\,, \, \, \, \,\,\,
{}^{(3)}{\nabla}_b u_a=\frac{1}{3} \Theta h_{ab} + \sigma_{ab}+\omega_{ab}\,,
\ee
where $a_a={u_a}^\dagger$ is the acceleration, $\Theta$ is the expansion, and the shear and vorticity are
$\sigma_{ab}$ and $\omega_{ab}$, respectively. Further, in the following, angle brackets applied to a vector
denote its projection onto tangent 3-spaces
\be
V_{\langle a\rangle} = h_a{}^b V_b\,.
\ee
When applied to a tensor, they denote the projected, anti-symmetric and trace free part 
\be 
W_{\langle ab \rangle}=[h_{(a}{}^c h_{b)}{}^d-\frac{1}{3}h^{cd}h_{ab}] W_{ab}\,.
\ee
The spatial curl of a variable is
\be 
({\rm curl} X)^{ab}=\epsilon^{cd \langle a}{}^{(3)}{\nabla}_c X^{b\rangle}{}_d\,,
\ee
where $\epsilon_{abc}=u^d\eta_{abcd}$ is the spatial volume.

Finally, we note that, since we treat the additional curvature as a fluid, we can write an energy
density and a pressure for this fluid, namely $\rho^R$ and $P^R$ \cite{Ananda:2008tx}.

\subsection{Linearized equations}

Fully non-linear governing equations valid in any spacetime (with suitable choice of $u_a$) 
can be found in Ref.~\cite{Carloni:2007yv}. In order to study cosmological perturbations, we linearize the 
equations around a Friedmann-Robertson-Walker background spacetime. The cosmological equations
for the background are
\begin{align}
&\Theta^2=3\tilde{\rho}^M+3\rho^R-\frac{{}^{(3)}{R}}{2}\,, \\
&{\Theta}^\dagger+\frac{1}{3}\Theta^2+\frac{1}{2}(\tilde{\rho}^M+3\tilde{P}^M)+\frac{1}{2}(\rho^R+3P^R)=0\,, \\
&{\rho}^M{}^\dagger+\Theta(\rho^M+P^M)=0\,. 
\end{align}

Linearization of the propagation and constraint equations gives\footnote{Note in the following that $E_{ab}$ and 
$H_{ab}$ are the electric and magnetic parts of the Weyl tensor: 
\be 
E_{ab}=C_{abcd}u^c u^d \,, \,\,\,\, H_{ab}=\frac{1}{2}C_{aecd}u^e\eta^{cd}{}_{bf}u^f\,.
\ee
}
\begin{align}
&{\Theta}^\dagger+\frac{1}{3}\Theta^2-{}^{(3)}\nabla^a a_a+\frac{1}{2}(\tilde{\rho}^M+3\tilde{P}^m)=
-\frac{1}{2}(\rho^R+3P^R)\,,\\
&{\omega}_a^\dagger+2 H \omega_a+\frac{1}{2}{\rm curl} a_a =0\,,\\
&{\sigma}_{ab}^\dagger+2H \sigma_{ab}+E_{ab}\,,
\end{align}
\be 
{E}_{ab}^\dagger+3HE_{ab}-{\rm curl}E_{ab}+\frac{1}{2}(\tilde{\rho}^M+\tilde{P}^M)\sigma_{ab}
=-\frac{1}{2}(\rho^R+P^R)\sigma_{ab}-\frac{1}{2}{\pi}^R{}^\dagger_{\langle ab \rangle}
-\frac{1}{2}{}^{(3)}\nabla_{\langle a}q^R{b\rangle}-\frac{1}{6}\Theta \pi_{ab}^R\,,
\ee
\begin{align}
&{H}_{ab}^\dagger+3H H_{ab}+{\rm curl}E_{ab}=\frac{1}{2}{\rm curl}\pi_{ab}^R\,,\\
& {}^{(3)}\nabla^b \sigma_{ab}-{\rm curl}\omega_a-\frac{2}{3}{}^{(3)}\nabla_a \Theta=-q_a^R\,,\\
&{\rm curl}\sigma_{ab}+{}^{(3)}\nabla_{\langle a}\omega_{b\rangle}-H_{ab}=0\,,
\end{align}
\begin{align}
&{}^{(3)}\nabla^b E_{ab}-\frac{1}{3}{}^{(3)}\nabla_a\tilde{\rho}^M=-\frac{1}{2}{}^{(3)}\nabla^b \pi_{ab}^R
+\frac{1}{3}{}^{(3)}\nabla \rho^R-\frac{1}{3}\Theta q_a^R\,,\\
&{}^{(3)}\nabla^b H_{ab}-(\tilde{\rho}^M+\tilde{P}^M)\omega_a=-\frac{1}{2}{\rm curl}q_a^R+(\rho^R+P^R)\omega_a\,,
\end{align}
\be 
{}^{(3)}\nabla^a\omega_a=0\,.
\ee
And the linearized conservation equations are
\begin{align}
 {\rho}^M{}^\dagger&=-\Theta(\rho^M+P^M)\,,\\
{}^{(3)}\nabla^a P^M&=-(\rho^M+P^M){u}^a{}^\dagger\,,
\end{align}
\begin{align}
&{\rho}^R{}^\dagger+{}^{(3)}\nabla^a q_a^R=-\Theta(\rho^R+P^R)+\rho^M\frac{F'}{F^2}{R}^\dagger\,,\\
&{q}^R{}^\dagger_{\langle a \rangle}+{}^{(3)}\nabla_a P^R+{}^{(3)}\nabla^b \pi_{ab}^R
=-\frac{4}{3}\Theta q_a^R-(\rho^R+P^R){u}_a^\dagger+\rho^M\frac{F'}{F^2}{}^{(3)}\nabla_a R\,.
\end{align}

\subsection{Scalar equations}

In order to study the linearized dynamics, we define the covariant gauge invariant quantities\footnote{ 
We should note that the term `gauge invariant' is used to mean different things in perturbation theories. 
In metric perturbation theory, we refer to a quantity as being gauge invariant if it does not change under a 
gauge transformation. However, a stronger notion of gauge invariance is introduced through the Stewart-Walker
lemma \cite{Stewart:1974uz},
 that an inhomogeneous linearly perturbed quantity is gauge invariant if the quantity vanishes in the background.
Such a quantity is often referred to as identification gauge invariant. See Ref.~\cite{Malik:2012dr} for more
discussion on this point.}
\be 
\mathcal{D}^M_a=\frac{a}{\rho^M}{}^{(3)}\nabla_a\rho^M\,, \hspace{1cm}
Z_a=a{}^{(3)}\nabla_a\Theta\,, \hspace{1cm}
C_a=a{}^{(3)}\nabla_a{}^{(3)} R\,,
\ee
as well as the gradients describing inhomogeneities in the Ricci scalar
\be 
\mathcal{R}_a=a{}^{(3)}\nabla_a R\,, \hspace{1cm}
\mathfrak{R}_a=a{}^{(3)}\nabla_a{R}^\dagger\,.
\ee
Dynamical and constraint equations for these variables can be found in Ref.~\cite{Carloni:2007yv}. However, 
since we want to consider scalar perturbations that govern the formation of structure in the universe, we need
to use scalar variables. These are obtained by using a local decomposition. The variables of interest are then obtained by applying
${}^{(3)}\nabla^a$ to those definitions above to give
\be 
\Delta_M=a{}^{(3)}\nabla^a\mathcal{D}_a^M\,, \hspace{7mm}
Z=a{}^{(3)}\nabla^aZ_a\,, \hspace{7mm}
C=a{}^{(3)}\nabla^aC_a\,, \hspace{7mm}
\mathcal{R}=a{}^{(3)}\nabla^a \mathcal{R}_a \,, \hspace{7mm}
\mathfrak{R}=a{}^{(3)}\nabla^a \mathfrak{R}_a \,.
\ee
Then, assuming the matter content to be well described by a barotropic fluid with equation of state $P^M=w\rho^M$, the evolution equations for the variables are
\begin{align}
\label{eq:cov1}
&{\Delta}_M^\dagger=w\Theta\Delta_M-(1+w)Z\,,\\
&{Z}^\dagger=\Bigg[\frac{R^\dagger F'}{F}-\frac{2\Theta}{3}\Bigg]Z+\Bigg[\frac{3(w-1)(3w+2)}{6(w+1)}\tilde{\rho}^M
+\frac{2w\Theta^2+3w(\rho^R+3P^R)}{6(w+1)}\Bigg]\Delta_M+\frac{\Theta F'}{F}\mathfrak{R}\nn\\
&+\Bigg[\frac{1}{2}-\frac{1}{2}\frac{fF'}{F^2}-\frac{F'}{F}\tilde{\rho}^M+{R}^\dagger\Theta\Big(\frac{F'}{F}\Big)^2
+{R}^\dagger\Theta\frac{F''}{F}\Bigg]\mathcal{R}-\frac{w}{w+1}{}^{(3)}\nabla^2\Delta_M
-\frac{F'}{F}{}^{(3)}\nabla^2\mathcal{R}\,,
\label{eq:cov2}
\end{align}

\begin{align}
\label{eq:cov3}
&{\mathcal{R}}^\dagger=\mathfrak{R}-\frac{w}{w+1}{R}^\dagger\Delta_M\,,\\
&{\mathfrak{R}}^\dagger=-\Bigg(\Theta+2{R}^\dagger\frac{F''}{F'}\Bigg)\mathfrak{R}-{R}^\dagger Z
-\Bigg[\frac{(3w-1)}{3}\frac{\rho^M}{F'}+3\frac{w}{w+1}(P^R+\rho^R)\frac{F}{F'}
+\frac{w}{3(w+1)}{R}^\dagger\Bigg(\Theta-3{R}^\dagger\frac{F''}{F'}\Bigg)\Bigg]\Delta_M\nn\\
\label{eq:cov4}
&-\Bigg[\frac{1}{3}\frac{F}{F'}+\frac{F'''}{F}{R^\dagger}^2+\Theta\frac{F''}{F}{R}^\dagger-\frac{2}{9}\Theta^2
+\frac{1}{3}(\rho^R+P^R)+{R}^{\dagger\dagger}\frac{F''}{F'}-\frac{1}{6}\frac{f}{F}+\frac{1}{2}(w+1)\tilde{\rho}^M
-\frac{1}{3}{R}^\dagger\Theta\frac{F'}{F}\Bigg]\mathcal{R}+{}^{(3)}\nabla^2\mathcal{R}\,,\\
&{C}^\dagger={}^{(3)}\nabla^2\Bigg[\frac{4wa^2\Theta}{3(w+1)}\Delta_M+2a^2\frac{F'}{F}\mathfrak{R}
-2a^2\frac{(\Theta F'-3{R}^\dagger F'')}{3F}\mathcal{R}\Bigg]\,,
\label{eq:cov5}
\end{align}
and a constraint equation
\be 
\label{eq:cov6}
\frac{C}{a^2}+\Bigg(\frac{4}{3}\Theta+\frac{2{R}^\dagger F'}{F}\Bigg)Z-2\tilde{\rho}^M\Delta_M
+\Bigg[2{R}^\dagger\Theta\frac{F''}{F}-\frac{F'}{F}\Big(f-2\rho^M+2{R}^\dagger\Theta F'\Big)\Bigg]\mathcal{R}
+\frac{2\Theta F'}{F}\mathfrak{R}-\frac{2F'}{F}{}^{(3)}\nabla^2\mathcal{R}=0\,.
\ee

\section{Relating the two approaches}
\label{sec:relation}

In the previous sections we have introduced cosmological perturbation theory using both the metric and 
covariant formalisms in $f(R)$ gravity. In this section we show how to relate one to the other focusing on the
covariant approach and showing how this maps to the metric approach.

First, the three-dimensional Ricci scalar is defined in the covariant approach as
\be 
^{(3)}R=R+2R_{bd}u^bu^d-\frac{6}{a^2}\H^2\,.
\ee
To compare, we split this into a homogeneous background and a perturbation as usual, 
${^{(3)}R=^{(3)}\bar{R}+\delta^{(3)}R}$. The background
is zero, $^{(3)}\bar{R}=0$, for a flat FLRW spacetime. The perturbation, from Ref.~\cite{Bruni1992}, can be written
in terms of metric perturbation variables as
\be 
\delta^{(3)}R=\frac{4}{a^2}\nabla^2\Big[\psi-\H(v+B)\Big]\,.
\ee
Using metric perturbation theory, we can calculate the 3-Ricci scalar to obtain \cite{Malik2001} 
\be
{\delta^{(3)}R}=\frac{4}{a^2}\nabla^2\psi\,.
\ee
The definition of the curvature perturbation in the comoving gauge in terms of variables in an arbitrary gauge is
\be 
\psi_{\rm com}=\psi-\H(v+B)\,,
\ee
from which we can see that the curvature quantities in the covariant approach are equivalent to the quantities in the metric
approach in the comoving gauge.
That is,
\be 
\delta^{(3)}R=\frac{4}{a^2}\nabla^2 \psi_{\rm com}\,,
\ee
This can be written in terms of longitudinal gauge quantities, or Bardeen variables, as
\be 
\delta^{(3)}R=\frac{4}{a^2}\nabla^2 (\Psi - \H v_\ell)\,.
\ee
The four dimensional Ricci scalar is derived above in Section~\ref{sec:metric} and is given
in the comoving gauge in terms of Bardeen variables as
\be 
\delta R_{\rm com}[\ell]=\frac{2}{a^2}\nabla^2(2\Psi-\Phi)-\frac{6}{a^2}\Big[\H(\dot{\Phi}+3\dot{\Psi})
+\ddot{\Psi}+2(\H^2+\dot{H})\Phi
+(2\H^3-\ddot{\H})v_\ell\Big]
\ee

Now we consider kinematical quantities, starting with the expansion scalar, 
in the covariant approach defined as $\Theta=\nabla_a u^a$. This can then
be split into a homogeneous background and a linear perturbation as
\be 
\Theta = \bar{\Theta}+\delta\Theta\,.
\ee
The background expansion is $\bar{\Theta}=3\H / a$. Using the definition of $u^M_a$ in terms of metric 
perturbation theory we arrive at
\be 
\label{eq:thetacov}
\delta\Theta = -\frac{3}{a}\Big[\H \phi+\dot{\psi}+\frac{1}{3}\nabla^2(v-\dot{E})\Big]\,.
\ee

One difference between the two formalisms is in the assumed time-like vector field with which to describe the
spacetime.
The covariant approach assumes a four-velocity, taken to be comoving with the matter, $u_M^a$, 
while the metric formalism assumes the FLRW metric as a background. In the latter, the fundamental vector
field, $n^a$ is orthogonal to constant$-\eta$ hypersurfaces, and has components
\begin{align}
n^a&=\frac{1}{a}(1-\phi,-B_,{}^i+S^i)\,,\\
n_a&=-a(1+\phi,0)\,.
\end{align}
Thus, in metric perturbation theory the expansion scalar is 
\be 
\widetilde{\delta\Theta}=-\frac{3}{a}\Big[\H \phi+\dot{\psi}+\frac{1}{3}\nabla^2(B-\dot{E})\Big]\,.
\ee
This is not such a problem, it arises simply because in metric perturbation theory we have a choice of 
unit timelike vector field.
If we want to compare the two approaches we can simply evaluate both in the comoving gauge, for which
$v=B=0$. So, Eq.~(\ref{eq:thetacov}), becomes
\be 
\delta\Theta= -\frac{3}{a}\Big[\H \phi_{\rm com}+\dot{\psi}_{\rm com}
-\frac{1}{3}\nabla^2 \dot{E}_{\rm com}\Big]\,,
\ee
and on using the relationships between the comoving gauge variables and the longitudinal gauge (or Bardeen) variables,
\begin{align}
\phi_{\rm com}&=\Phi+\H v_{\ell}+\dot{v_\ell}\,,\\
\psi_{\rm com}&=\Psi-\H v_\ell\,,\\
\dot{E}_{\rm com}&=v_\ell\,,
\end{align}
we obtain
\be 
\delta\Theta= -\frac{3}{a}\Big[\H \Phi+\dot{\Psi}
+(\H^2-\dot{\H})v_\ell-\frac{1}{3}\nabla^2 v_{\ell}\Big]\,.
\ee
Similarly, the acceleration, $a_a=u_{a;b}u^b$ in the comoving gauge is
\be 
a_i=\phi_{{\rm com},i}\,,
\ee
which, in terms of Bardeen variables, is
\be 
a_i=\Phi_{,i}+\H v_{\ell,i}+\dot{v}_{\ell,i}\,.
\ee

\subsection{Gauge invariant covariant quantities}

Now, we show how to relate the gauge invariant gradients defined above to metric perturbation quantities. 
Again, as above, we work in the comoving gauge. In this gauge, the projected covariant derivative, defined as
$a{}^{(3)}\nabla_i=h^b{}_i\nabla_b$ is simply the covariant derivative on the spatial hypersurfaces and, since
we are working with a flat background, is simply a partial derivative: $a{}^{(3)}\nabla_i=\partial_i$. Thus, we obtain
\begin{align}
\mathcal{D}^M_i &= \delta_{{\rm com},i}\,,\\
{Z}_i &= -\frac{3}{a}\Big[\H\Phi_{,i}+\dot{\Psi}_{,i}+(\H^2-\dot{\H})v_{\ell,i}-\frac{1}{3}\nabla^2v_{\ell,i}\Big]\,,\\
C_i &= \frac{4}{a^2}\nabla^2(\Psi_{,i}-\H v_{\ell ,i})\,,\\
\mathcal{R}_i &= \partial_i\delta R_{\rm com}[\ell]\,,\\
\mathfrak{R}_i &= \frac{1}{a}\dot{\delta R}_{\rm com}[\ell]-\frac{1}{a}(\Phi_{,i}+\H v_{\ell,i}+\dot{v}_{\ell,i})\dot{\bar{R}}\,.
\end{align}
Note that we have left the variables in terms of $\delta R_{\rm com}[\ell]$, since we can then simply substitute
this into the covariant equations later which will then give us equations comparable to the longitudinal gauge
metric equations presented in Section~\ref{sec:long}. The scalar gauge invariant covariant quantities are 
then related to metric variables through
\begin{align}
\Delta_M &=\nabla^2\delta_{\rm com}\,,\\
Z &= -\frac{3}{a}\nabla^2\Big[\H\Phi+\dot{\Psi}+(\H^2-\dot{\H})v_\ell-\frac{1}{3}\nabla^2v_\ell\Big]\,,\\
C &= \frac{4}{a^2}\nabla^4(\Psi-\H v_\ell)\,,\\
\mathcal{R} &= \nabla^2\delta R_{\rm com}[\ell]\,,\\
\mathfrak{R} &= \frac{1}{a}\nabla^2\dot{\delta R}_{\rm com}[\ell]+\frac{1}{a}\nabla^2(\Phi+\H v_\ell+\dot{v}_\ell)\dot{\bar{R}}\,.
\end{align}

\subsection{Equations}

Having now presented the gauge invariant covariant variables in terms of metric perturbation variables, we nowshow how to convert from the equations in the covariant approach to those in the metric approach. We will use the case of general relativity, for which $fR)=R$, to highlight the procedure. We first note that
a dagger derivative applied to a perturbed quantity in the comoving gauge is
\be 
x^\dagger = \frac{1}{a}\dot{x}\,.
\ee
The equivalency is best shown by first performing a harmonic decomposition, such that
\be 
^{(3)}\nabla^2Q=-\frac{k^2}{a^2}Q^{(k)}\,.
\ee
This removes the $^{(3)}\nabla^2$ from the equations in the covariant approach, thus allowing a more direct comparison with the metric approach. Then, the set of equations governing the scalar variables can be written as two, second order differential equations:
 \begin{align}
 \Delta_M^{(k)\dagger\dagger}+\Bigg[\Big(\frac{2}{3}-w\Big)\Theta-\frac{R^{\dagger}F'}{F}\Bigg]\Delta_M^{(k)\dagger}-\Bigg[w\frac{k^2}{a^2}-w(3P^R+\rho^R)-\frac{2wR^\dagger \Theta F'}{F}-\frac{(3w^2-1)\rho^M}{F}\Bigg]\Delta_M^{(k)}\nonumber\\
 =\frac{1}{2}(w+1)\Bigg[2\frac{k^2}{a^2}F'-1+\Big(f-2\rho^M+2R^\dagger \Theta F'\Big)\frac{F'}{F^2}-2R^\dagger\Theta\frac{F''}{F}\Bigg]{\mathcal R}^{(k)}-\frac{(w+1)\Theta F'}{F}{\mathcal R}^{(k)\dagger}\,,
 \end{align}
 \begin{align}
 F'{\mathcal R}^{(k)\dagger\dagger}+\Big(\Theta F'+2R^\dagger F''\Big){\mathcal R}^{(k)\dagger}-\Bigg[\frac{k^2}{a^2}F'+\frac{2}{9}\Theta^2 F'-(w+1)\frac{\rho^M}{2F}F'-\frac{1}{6}\Big(\rho^R+3P^R\Big)F'\nonumber\\
 -\frac{F}{3}+\frac{f}{6F}F'+R^\dagger\Theta\frac{F'^2}{6F}-R^{\dagger\dagger}F''-\Theta F'' R^\dagger - F^{(3)}R^{\dagger 2}\Bigg]{\mathcal R}^{(k)}=-\Bigg[\frac{1}{3}(3w-1)\rho^M \nonumber\\
 +\frac{w}{w+1}\Bigg(F'' (R^\dagger)^2+(\rho^R+P^R)F+\frac{7}{3}R^\dagger\Theta F' +R^{\dagger\dagger}F'\Bigg)\Bigg]\Delta_M^{(k)}-\frac{(w-1)R^\dagger F'}{w+1}\Delta_M^{(k)\dagger}\,.
 \end{align}

On taking the general relativistic limit, the equation governing the evolution of the energy density perturbation for a flat FLRW universe dominated by dust, is
\be 
\Delta_M^{(k)\dagger\dagger}+\frac{2}{3}\bar{\Theta}\Delta_M^{(k)\dagger}-\frac{1}{2}\kappa\rho^{M}\Delta_M^{(k)}=0\,.
\ee
Using the relationships between the covariant and metric perturbation quantities in the comoving gauge, this can be re-written as 
\be 
\ddot{\delta}_{\rm com}+\H\dot{\delta}_{\rm com}-4\pi G \rho a^2 \delta_{\rm com} =0 \,.
\ee
This is the usual equation for the evolution of the density contrast in the comoving gauge, which verifies the transformation between the two approaches. 

\section{Discussion}
\label{sec:dis}

In this article we have provided, for the first time, a method for relating the most popular
two methods of modeling cosmological perturbations -- the covariant and metric approaches -- 
to one another for $f(R)$ gravity. This builds upon work presented in Ref.~\cite{Bruni1992} for standard Einstein gravity.
We started by reviewing $f(R)$ gravity, and then both the metric and covariant approaches to cosmological
perturbations. We presented the governing equations for scalar perturbations 
in the metric approach in both the longitudinal gauge and the
uniform curvature gauge, as well as presenting the equations in Bardeen's variables (which amounts to using
the longitudinal gauge, but with the comoving density contrast). The governing equations in the covariant
approach were then presented, again for scalar perturbations, in terms of the covariant gauge invariant quantities.

Then, in Section \ref{sec:relation}, we presented the relationship between the variables in the covariant and
metric approaches. For the curvature variables such as the 3-dimensional Ricci scalar, the covariant 
variables are essentially already in a form equivalent to the comoving gauge of metric perturbation theory. For 
kinematic variables, such as the expansion scalar, the covariant variables are equivalent to the metric perturbation
theory variables only in the comoving gauge, due to the choice of the velocity four vector $u_M^a$ 
as opposed to the unit timelike vector $n^a$, which depends on metric perturbations.
Having presented this relationship, we then outlined the method in which one can transfer from the covariant
equations to the metric equations, and vice versa. 

It is not surprising that this relationship exists, since the two approaches are complementary methods with 
which to describe inhomogeneities on top of a homogeneous and isotropic FLRW background, with the covariant approach mapping to the comoving gauge \cite{Malik:2012dr}. However, since the two
approaches are different, there will naturally be problems that one or other of the methods are more suitable 
to solve. This paper allows one to compare calculations done in one of the approaches to the other, thus enabling
a deeper understanding of the predictions made by different $f(R)$ cosmological theories.

\section*{Acknowledgements}
AJC is grateful to Peter Dunsby for discussions that motivated this project, to Karim Malik for helpful comments on an earlier draft, and to Megan Reavis for careful reading of the manuscript. He is also grateful to Iain Brown, and Reza Tavakol for discussions and to the ICN and IA, 
Universidad Nacional Aut\' omata de M\' exico (UNAM), and in particular Carlos Hidalgo and Cesar Lopez-Monsalvo, for generous hospitality while part of this work was completed. This work is supported in part by the U.S. Department of Energy under Grant No. DE-FG02-97ER41029 at the University of Florida.

\begin{appendix}

\section{Notation}
\label{app:notation}

In this paper we follow the notation of Ref.~\cite{MW2008}, however in this Appendix we show how to relate notation
to that of Ref.~\cite{Bruni1992}, where the perturbed line element is
\be 
ds^2=a^2(\eta)\Big[-(1+2A)d\eta^2-2B_{\alpha}dx^\alpha d\eta
+\Big((1+2H_L)\gamma_{\alpha\beta}+2H_{T|\alpha\beta}\Big)dx^\alpha dx^\beta\Big]\,,
\ee
where here Greek indices run over the spatial coordinates. This is also discussed in Ref.~\cite{Carrilho:2015cma}.
The perturbations are then decomposed as
\begin{align}
B_{\alpha}&=B_{|\alpha}+B^S_\alpha\,,\\
H_{T\alpha\beta}&=\nabla_{\alpha\beta}H_T+H^S_{T(\alpha|\beta)}+H^{TT}_{T\alpha\beta}\,,
\end{align}
where
\be 
\nabla_{\alpha\beta} \zeta = \zeta_{|\alpha\beta}-\frac{1}{3}\nabla^2\zeta\,,
\ee
for some scalar, $\zeta$. Thus, we arrive at the equivalences between the scalar perturbations in the conventions of 
Malik and Wands (left) and Bruni et al (right):
\begin{align}
&\phi \longleftrightarrow A \\
&B \longleftrightarrow -B \\
&\psi \longleftrightarrow \frac{1}{3}\nabla^2H_T- H_L \\
&E \longleftrightarrow H_T\,.
\end{align}

\end{appendix}

\bibliography{fR_apr16.bbl}

\end{document}